\documentclass{article}
\usepackage{frascatiphys,here,graphicx,subfigure}
\usepackage{amssymb}
\begin{document}
\title{ 
Heavy ion collisions phenomenology overview.}
\author{
Carlos A. Salgado\footnote{carlos.salgado@cern.ch}        \\
{\em Dipartimento di Fisica, Universit\`a di Roma "La Sapienza"}\\
{\em and INFN, Roma, Italy}\\
{\em and}\\
{\em Departamento de F\'\i sica de Part\'\i culas and IGFAE,}\\
{\em Universidade de Santiago de Compostela, Spain} \\
}
\maketitle
\baselineskip=11.6pt
\begin{abstract}
The reach of collider energies in heavy-ion collisions has profoundly changed our understanding of QCD under extreme conditions. I review some these new developments and comment on the properties of the produced medium as extracted from experimental data, as well as the exciting new opportunities which will be open at the LHC.
\end{abstract}
\baselineskip=14pt

QCD is a theory with a very rich dynamical structure but difficult to solve in many situations of phenomenological interest. This structure includes confinement and chiral symmetry breaking as main vacuum properties, a complex phase diagram and hadronic spectrum, asymptotic freedom and others. Among these, only asymptotic freedom has allowed to make extensive experimental tests of the precision of the theory in the short distance regime of the interaction. Lattice calculations allows to test the long distance dynamics giving excellent results for static quantities \cite{davies}, but with limitations to study out of equilibrium situations. Most of the present phenomenological applications require, however,  this real-time dynamics. Two examples arise: the recent interpretations about the structure of resonances on different mass regions of the spectrum -- extensively discussed at this conference -- or the transport properties of the hot medium created in nuclear collisions. The common question of both topics could be phrased as: {\it what are the relevant building blocks in situations where collective behavior appears and how they organize?} 

The experiments of heavy-ion collisions at high energy attempt to answer this question for the hot part of the phase space diagram. The dynamical properties of the created matter, as the equation of state or different transport coefficients, are accessible experimentally and the findings are being interpreted theoretically. Several questions can be addressed which are normally categorized depending on the time scale as i) initial state of the system, ii) thermalization and evolution, iii) probes of the medium. We follow this classification in the following.

\section{The initial state and the Color Glass Condensate}

The relevant part of the colliding nuclei (or hadrons in general) wave function is dominated at high energies by Lorentz-boosted short-living quantum fluctuations which, with several degrees of sophistication, can be computed perturbatively once some initial  condition is provided. This 'small-$x$ gluons' are produced by sequential splitting in a branching process which makes its number to grow exponentially in rapidity $y=-\log x$, the variable playing the role of time for the evolution. When the density number of gluons is very high, the probability of fusion begins to compensate that of branching and a phenomenon of saturation appears \cite{Gribov:1984tu} -- the corresponding scale when this happens is called the saturation scale $Q^2_{\rm sat}$. 

A successful implementation of this physics is known under the the generic name of {\it Color Glass Condensate} \cite{Iancu:2003xm}. It provides a general framework for the whole collision, based on an effective theory separating the fast modes in the wave function from the {\it generated} slow modes, associated to small-$x$ gluons, which are treated as classical fields. The quantum evolution equation of this setup is also known and, remarkably, recent attempts exist aiming to provide the link to the subsequent evolution into a thermal system \cite{Lappi:2006fp}. 

\begin{figure}
\begin{minipage}{0.5\textwidth}
\begin{center}
\includegraphics[width=\textwidth]{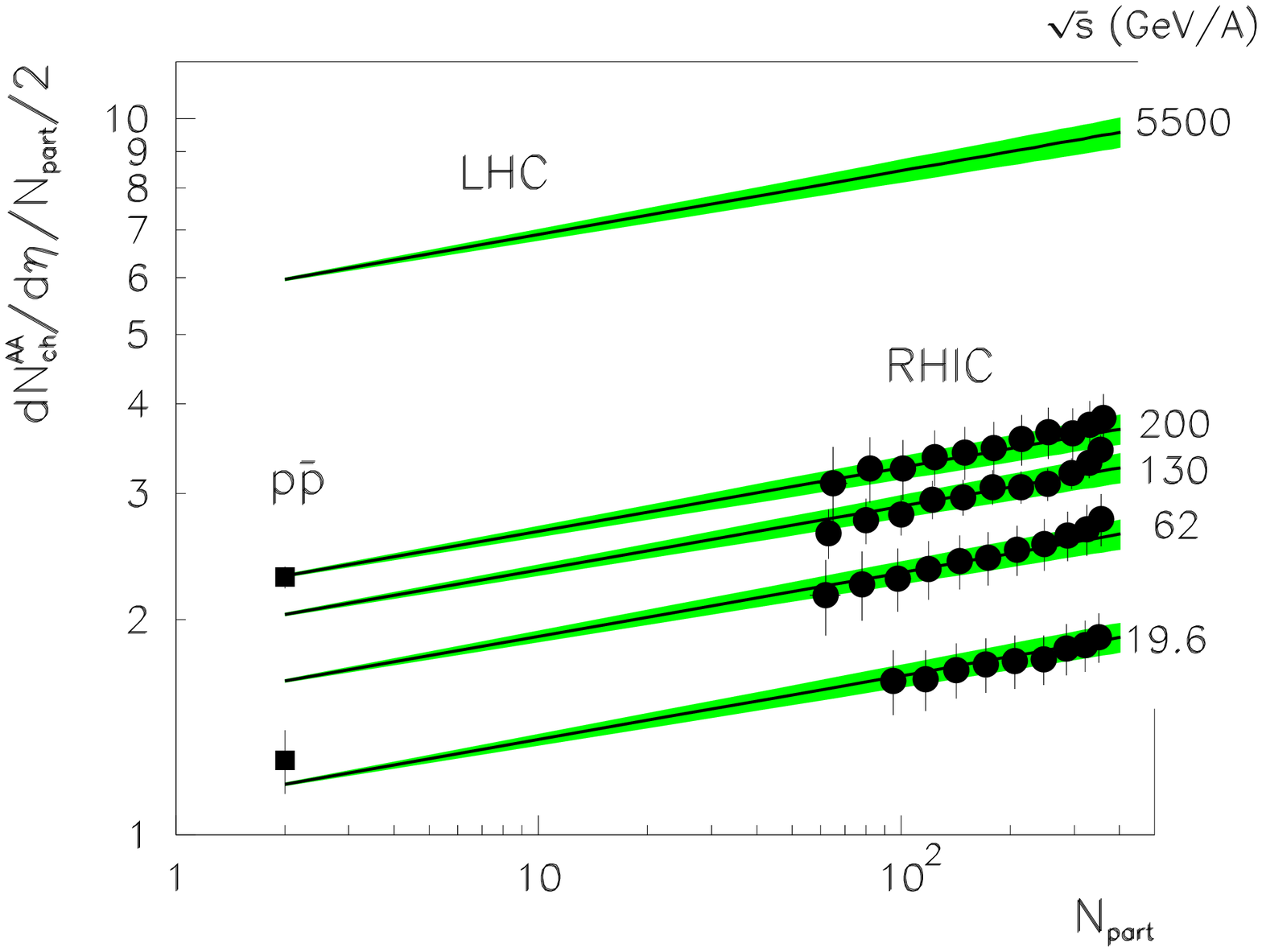}
\end{center}
\end{minipage}
\hfill
\begin{minipage}{0.5\textwidth}
\begin{center}
\includegraphics[width=0.8\textwidth]{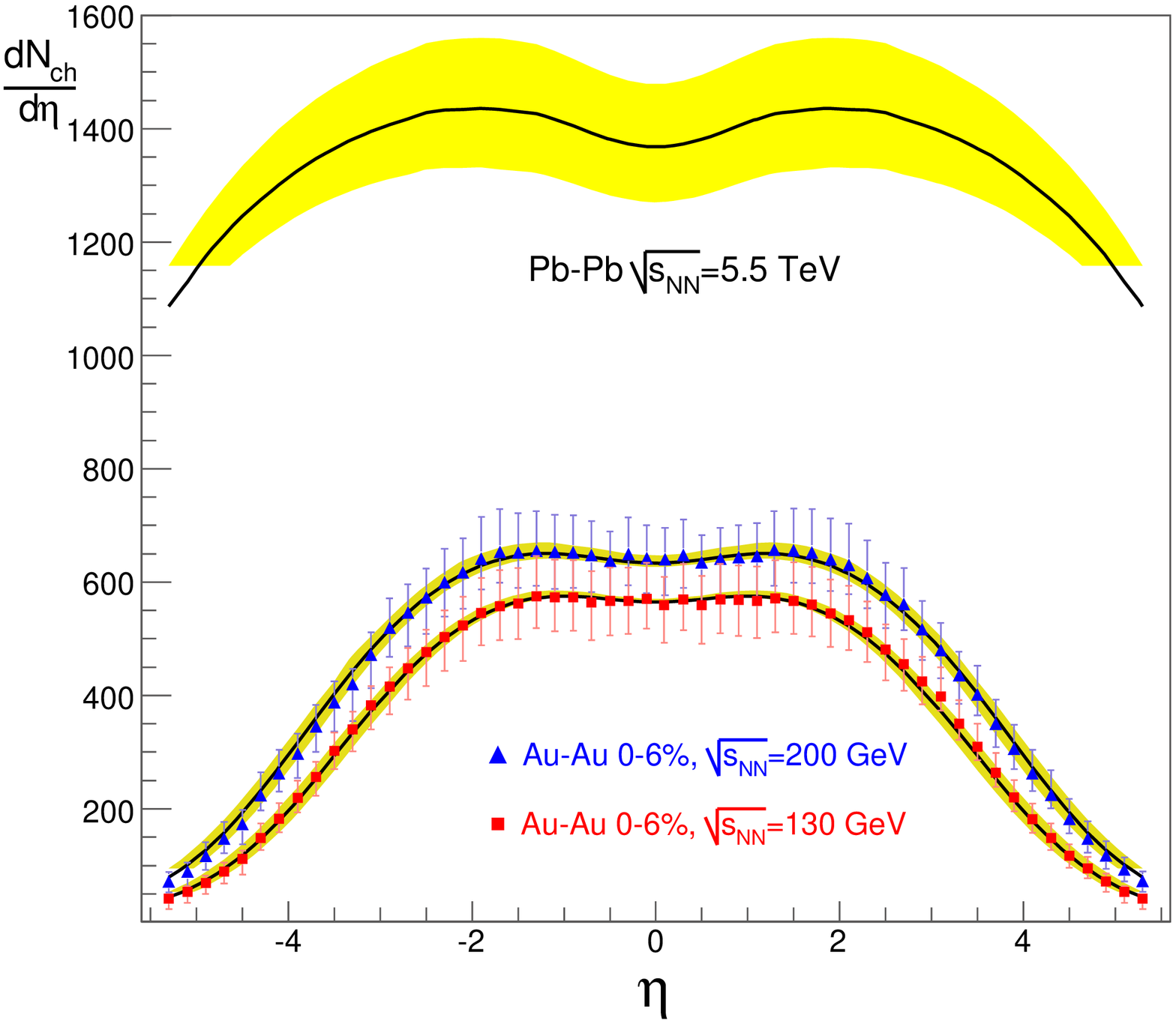}
\end{center}
\end{minipage}
\caption{Left: Particle multiplicities in nuclear collisions at different energies using the simple parametrization (\protect\ref{eqmult}) \protect\cite{Armesto:2004ud}. Right: Rapidity dependence of the multiplicities in central nucleus-nucleus collisions from Ref. \cite{Albacete:2007hg}.}
\label{fig1}
\end{figure}

Interestingly, this formalism provides a way of computing multiparticle production. In its simplest implementation, the total multiplicity is proportional to the saturation scale times a geometric factor \cite{Kharzeev:2000ph}. A particularly economic description is given by the pocket formula\cite{Armesto:2004ud}
\begin{equation}
\frac{2}{N_{\rm part}}
\frac{dN^{AA}}{d\eta}\Bigg\vert_{\eta\sim 0}=N_0\sqrt{s}^\lambda
N_{\rm part}^{\frac{1-\delta}{3\delta}}\, ,
\label{eqmult}
\end{equation}
where the energy and system size dependences ($\lambda=0.288$ and $\delta=0.79$) come from fits to DIS data and only a total normalization factor $N_0$=0.47 is introduced. Fig. \ref{fig1} shows the comparison of this simple formula with available data \cite{Back:2004je}. A step forward in this  phenomenology is the description of the experimental multiplicity data by the CGC evolution equations including part of the NLO corrections  \cite{Albacete:2007hg} -- also plotted in Fig. \ref{fig1}.

\section{The soft bulk and the hydrodynamical evolution}

The evolution of the transient system formed in a heavy ion collision should follow a hydrodynamical behavior if thermal equilibrium is reached. In this case, the hydrodynamical equations give the evolution of flow fields, densities and pressures for a given initial configuration provided the equation of state of the system is known. The signals from this behavior are one of the most direct probes of the degree of thermalization in heavy-ion collisions  -- see e.g. \cite{Hirano:2007gc} for a recent summary. 

A particularly important measurement is the azimuthal anisotropy with respect to the reaction plane for non-central nuclear collisions\footnote{The extension of the nuclei allow for a definition of the relative distance of the centers at every collision -- the impact parameter -- so that different system geometries and densities can be studied.}: for these collisions, the interaction region is asymmetric with different gradients of density depending on the azimuthal angle. In a hydrodynamical medium, these gradients lead to different accelerations of the particles in the medium, so that the spatial anisotropy translates into momentum anisotropies. This effect is normally parametrized by the first non-trivial coefficient in the Fourier expansion, $v_2$, which, for the hydrodynamical interpretation, is called {\it elliptic flow} \cite{Ollitrault:1992bk}.

The experimentally measured anisotropy is in agreement with a hydrodynamical description with negligible viscosity. This has two important consequences, on the one hand, it indicates that the medium is in a local thermal equilibrium during the evolution and, on the other hand, it characterizes the medium as a liquid rather than a gas -- which would present a large viscosity.

\section{Hard processes as probes of the medium}

Hard processes are those involving large momentum exchanges, for which the factorization theorems of QCD allow for a separation between short- and long-distance contributions to the cross section
\begin{equation}
\sigma^{AB\to h}=
f_A(x_1,Q^2)\otimes f_B(x_2,Q^2)\otimes \sigma(x_1,x_2,Q^2)\otimes D_{i\to h}
(z,Q^2)\, .
\label{eqhard}
\end{equation}
Here, the short-distance perturbative cross section, $\sigma(x_1,x_2,Q^2)$ takes place in a very short time, $1/Q$, so that it is unmodified in nuclear collisions. The long-distance terms are non-perturbative quantities involving scales ${\cal O}(\Lambda_{\rm QCD})$ which are modified by the interaction with the medium. These modifications allow to characterize the medium properties -- see e.g. \cite{Salgado:2007rs} for a recent review.

A conceptually simple example is the $J/\Psi$, whose production cross section can be written as
\begin{equation}
\sigma^{hh\to J/\Psi}=
 f_i(x_1,Q^2)\otimes f_j(x_2,Q^2)\otimes
\sigma^{ij\to [c\bar c]}(x_1,x_2,Q^2)
 \langle {\cal O}([c\bar c]\to J/\Psi)\rangle\, ,
\end{equation}
where now $ \langle {\cal O}([c\bar c]\to J/\Psi)\rangle$ describes the hadronization of a $c\bar c$ pair in a given state (for example a color octet) into a final $J/\Psi$. In the case that the pair is produced inside a hot medium this long-distance part is modified: the potential between the pair is screened and the hadron is dissolved, making $ \langle {\cal O}([c\bar c]\to J/\Psi)\rangle\to 0$. The experimental observation of this effect is a disappearance of the $J/\Psi$ in nuclear collisions \cite{Matsui:1986dk}. This suppression has been discovered in experiments at the CERN SPS \cite{jpsiSPS} and measured also at RHIC \cite{jpsiRHIC}.

The $J/\Psi$-suppression involves the modification of the non-perturbative hadronization probability.
From the computational point of view, a theoretically simpler case is the modification of the {\it evolution} of fragmentation functions of high-$p_t$ particles due to the presence of a dense or finite--temperature medium. Here, highly energetic partons, produced in a hard process, propagate through the produced matter, loosing energy by medium-induced gluon radiation -- see Sec. \ref{sec:highpt}.

\subsection{Nuclear parton distribution functions}

\begin{figure}[tbh]
\centering
\centering\includegraphics[width=0.7\textwidth]{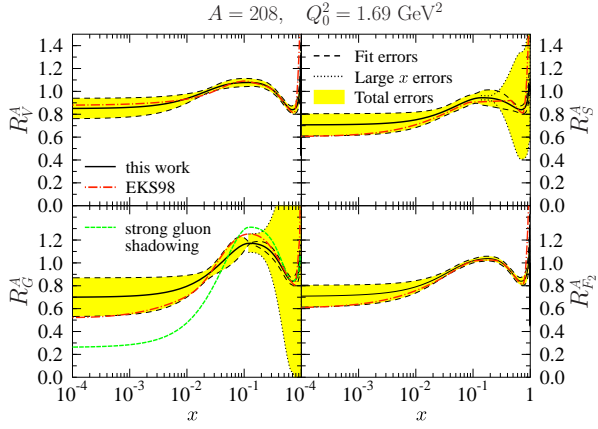}
\caption{Ratios of nuclear to free proton PDFs for different flavors at the initial scale $Q_0^2$=1.69 GeV$^2$ from \protect\cite{Eskola:2007my} with error estimates. The green line in the gluon panel is an attempt to check the strongest gluon shadowing supported by present data. }
\label{fig:nPDF}
\end{figure}

 A good knowledge of the PDFs is essential in any calculation of hard processes. The usual way of obtaining these distributions is by a global fit of data on different hard processes 
(mainly DIS) to obtain a set of parameters for the initial, non-perturbative, input $f(x,Q^2_0)$ to be evolved by DGLAP equations. Nuclear analyses (most recent ones in Refs. \cite{deFlorian:2003qf,Eskola:2007my,Hirai:2007sx}), using this procedure find a different initial condition, $f_A(x,Q^2_0)$, for the evolution which encodes the nuclear effects. Here, non-linear corrections to the evolution equations are usually neglected. An important consequence of these analysis is that present nuclear DIS and DY data can only constrain the distributions for $x\gtrsim 0.01$ -- see Fig. \ref{fig:nPDF}. By chance, this region covers most of the RHIC kinematics. For the LHC, where much smaller values of $x$ will be measured, a parallel proton-nucleus program will be essential as a benchmark for genuine hot-medium effects. 

\subsection{High-$p_t$ studies in heavy ion collisions: Jet quenching}
\label{sec:highpt}

Jet structures are expected to be modified when the branching process initiated after the perturbative production of a high-$p_T$ quark or gluon takes place into a thermal medium. The associated effects are generically known under the name of {\it jet quenching} and its simplest observational prediction is the suppression of the inclusive yields at high-$p_t$. This suppression can be traced back to a medium-modification of the fragmentation function $D_{i\to h}(z,Q^2)$ in Eq. (\ref{eqhard}).

\begin{figure}
\begin{minipage}{0.5\textwidth}
\begin{center}
\includegraphics[width=0.75\textwidth,angle=-90]{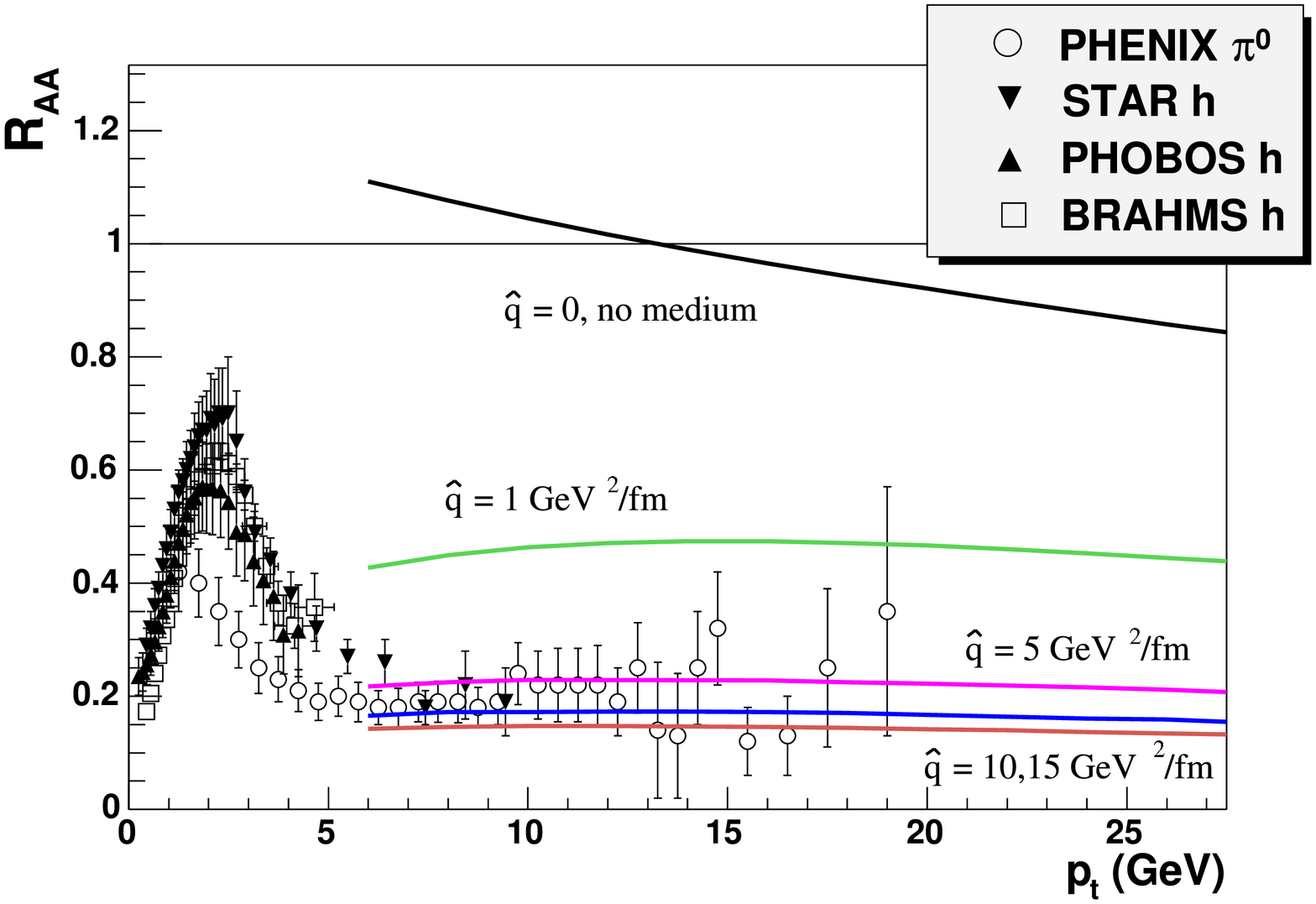}\\
\vskip 0.4cm
\centerline{  }
\end{center}
\end{minipage}
\hfill
\begin{minipage}{0.5\textwidth}
\begin{center}
\includegraphics[width=0.7\textwidth,angle=-90]{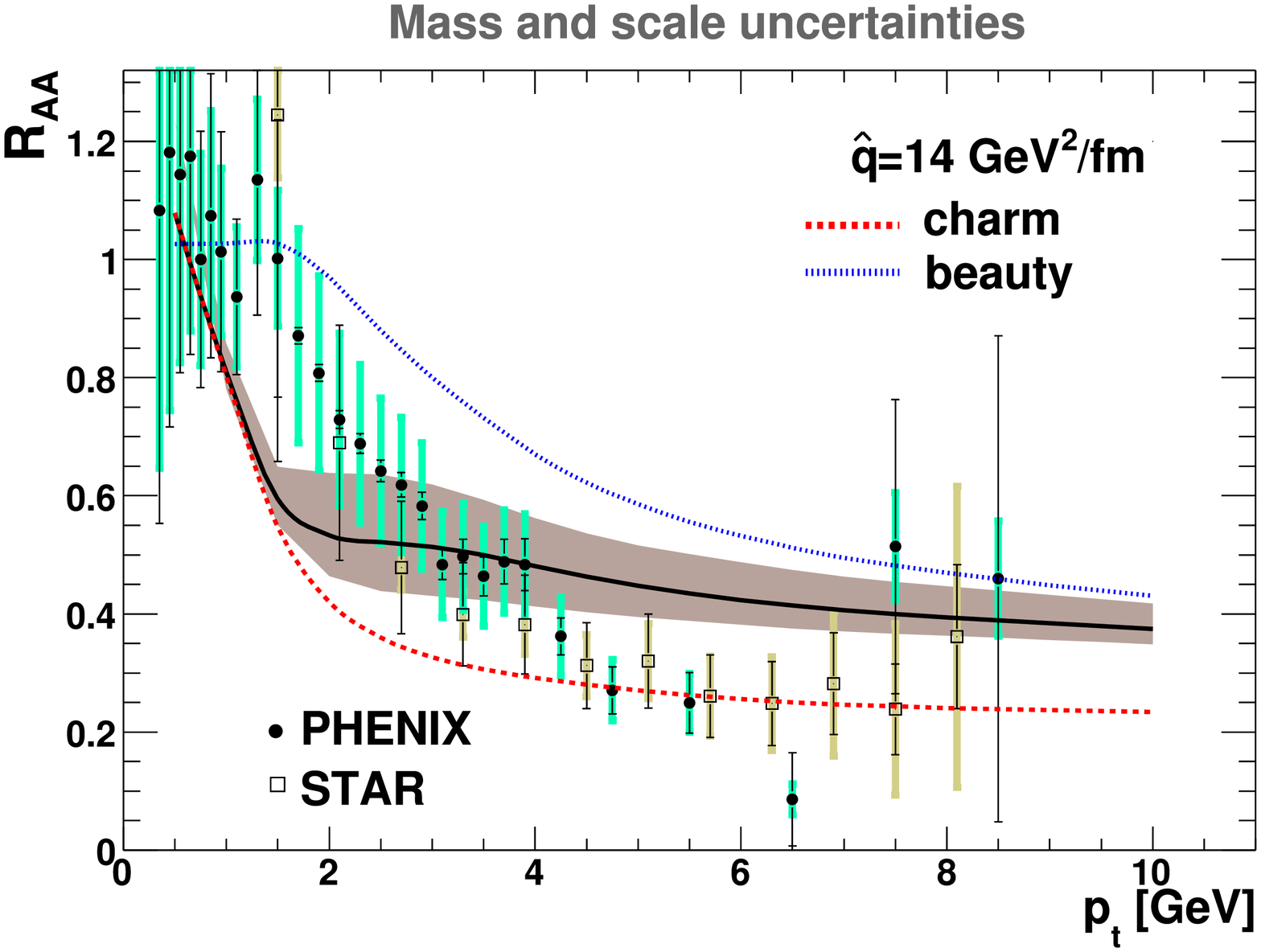}
\end{center}
\end{minipage}
\caption{Left: Nuclear modification factor, $R_{AA}$, for light hadrons in central
AuAu collisions \protect\cite{Eskola:2004cr}. Data from \protect\cite{Adcox:2001jp}. Right: $R_{AA}$ for non-photonic electrons with the corresponding uncertainty from the perturbative benchmark on the relative $b/c$ contribution \protect\cite{Armesto:2005mz}. Data from \cite{Abelev:2006db,Adare:2006hc}}
\label{fig:supp}
\end{figure}

A way of implementing these effects is by a redefinition of the splitting functions 
\begin{equation}
P^{\rm tot}(z)= P^{\rm vac}(z)+\Delta P(z,t),
\label{eq:medsplit}
\end{equation}
in the DGLAP evolution equations. Although this redefinition has not ben proved in general, it has been found to work assuming an independence of the multiple gluon emission when the rescattering with the nuclei is present\cite{Wang:2001if}. This possibility has been exploited in \cite{Armesto:2007dt} where the additional term in the splitting probability is just taken from the medium-induced gluon radiation \cite{Wang:2001if,Baier:1996sk,Salgado:2003gb} by comparing the leading contribution in the vacuum case.
\begin{equation}
\Delta P(z,t)\simeq \frac{2 \pi  t}{\alpha_s}\, 
\frac{dI^{\rm med}}{dzdt} ,
\label{medsplit}
\end{equation}
The fact that the medium-induced gluon radiation is finite in the soft and collinear limits allows for a simplification in which the medium-modified fragmentation functions are given by  
\begin{equation}
D_{i\to h}^{\rm med}(z,Q^2)=P_E(\epsilon)\otimes D_{i\to h}(z,Q^2)
\label{eqff}
\end{equation}
and where  $P_E(\epsilon)$ is given by a Poisson distribution\cite{Salgado:2003gb}. 
The medium-induced energy loss probability distribution $P_E(\epsilon)$ -- known as {\it quenching weights}, QW -- depends only on the in-medium path-length of the hard parton and the transport coefficient $\hat q$. The length is given by geometry and it is not a free parameter of the calculation -- although different geometries, including expansion, hydrodynamics, etc. could lead to slightly different results \cite{Renk:2006pk}. The transport coefficient encodes all the properties of the medium accessible by this probe and can be related to the average transverse momentum gained by the gluon per mean free path in the medium. Taking it as a free parameter of the calculation and fitting available data, a value of
\begin{equation}
\hat q=5....15\, {\rm GeV}^2/{\rm fm} 
\label{eq:qhat}
\end{equation}
is obtained \cite{Eskola:2004cr,Dainese:2004te}. Once this value is obtained, the formalism predicts the effects for other observables as heavy-quark suppression. In Fig. \ref{fig:supp} the description of the data for light mesons (used to fit the value of $\hat q$) and non-photonic electrons is shown. For the last, the uncertainty on the relative contribution from charm and beauty decays, shown by a band, is not yet under good theoretical control.  The description of the data within the formalism is reasonable but an experimental separation of both contribution will help to understand whether other effects \cite{Wicks:2007sn} are at work .

\subsection{Jets}

The most promising signal of the dynamics underlying jet quenching is the study of the modifications of the jet structures \cite{Salgado:2003rv} in which the characteristic angular dependence of the associated medium-induced radiation should be reflected. Experimentally, the main issue to overcome is the jet energy calibration in a high-multiplicity environment where small-$p_t$ cuts and more or less involved methods of background subtraction will be needed. From a theoretical point of view, identifying signals with small sensitivity to these subtractions is of primary importance \cite{Salgado:2003rv}. Due to these limitations, jet studies are not possible in AuAu collisions at RHIC but will be abundant at the LHC up to transverse energies of several hundred GeV. In the meantime, jet-like structures are being studied at RHIC by means of two- and three-particle correlations. 

An important step forward is the first measurement of two particle azimuthal correlations at large transverse momentum, with negligible combinatorial background \cite{Adams:2006yt}. These data support the picture of a very opaque medium with large energy losses, but with a broadening of the associated soft radiation hidden underneath the cut-off. Lowering this transverse momentum cut-off the different collaborations find non-trivial angular structures \cite{Adler:2005ee} in the form of a double-peak structure, in striking contrast with the typical Gaussian-like shape in proton-proton or peripheral AuAu collisions. Similar structures are found in large angle medium-induced radiation due to the LPM and Sudakov suppression of collinear gluons with energy $\omega\lesssim 2\hat q^{1/3}\sim 3$ GeV \cite{Polosa:2006hb}. In this framework, most of the energy is lost by radiation with negligible deposition in the medium. On the opposite limit, if a large fraction of the jet energy is deposited fast enough into a hydrodynamical medium it will be diffused by sound or dispersive modes. For very energetic particles, traveling faster than the speed of sound in the medium, a shock wave is produced with a characteristic angle which could also be at the origin of the measured structures \cite{conical}. Another interpretation of this effect is in terms of Cherenkov radiation \cite{cherenkov}. 

\section{Counting the valence quarks of exotic hadrons}

Very interesting effects appear in the intermediate region of $2\lesssim p_\perp\lesssim 6$ GeV/c. The most spectacular of them is the appearance of valence quark number scaling laws for baryons and mesons: (i) $R_{CP}$, the normalized ratio of high-$p_T$ yields in central to peripheral collisions, seems to depend only on the valence number of the produced particle; (ii) the elliptic flow parameter $v_2$ is universal when plotted as $v_2(p_\perp/n)/n$, $n$ being the number of valence quarks. A successful model to describe these features is a two component soft+hard model, in which the soft spectrum is assumed to come from the recombination of quarks in a medium in thermal equilibrium \cite{Friesvb}. The hard part of the spectrum is given by Eq. (\ref{eqhard}) with a simplified treatment of the energy loss. 

\begin{figure}[htb!]
\begin{center}
\includegraphics[scale=0.7]{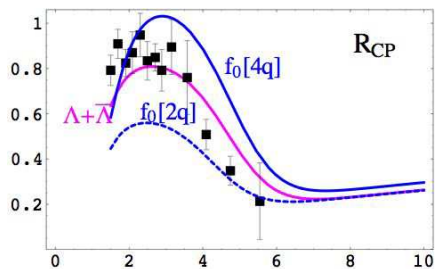}
\end{center}
\vspace{-0.5truecm}
\caption{${R_{CP}}$ for $\Lambda+\bar\Lambda$ and for $f_0(980)$ considered as a $q\bar q$ or a 4-quark meson.}
\label{fig:newrcp}
\end{figure}

In \cite{Maiani:2006ia} this model has been extended to the case of a 4-quark meson to study the sensitivity of these observables to make a case for the discovery of exotic states in heavy-ion collisions, in particular for the $f_0(980)$. In Fig. \ref{fig:newrcp},  $R_{CP}$ for $\Lambda+\bar\Lambda$ baryons is compared with experimental data from RHIC together with the model expectations for the corresponding effects on the $f_0(980)$ as being a normal meson or a 4-quark state -- the description for other hadrons could be found in  \cite{Maiani:2006ia}.
 
Although the analysis presented here is based on a given model implementation, the experimental facts on the quark counting rules indicate that the effect is more general and would survive more sophisticated implementations. These findings show that heavy-ion collisions are ideal tools to study the content of different resonances and to find a definitive answer to the structure of these exotic states. Clearly such a measurements are, at the same time, excellent playgrounds for the study of the relevant degrees of freedom of the produced medium. 

\section{Final comments}

Heavy-ion collisions, together with spectroscopy, have been the two most active areas of discoveries in the strong interaction in the last years. Both deal with the structure of extended objects for which first principle computations in QCD present some limitations. In these conditions a cross talk between theory and experiment is essential to make progress on the understanding of how macroscopic structures organize in QCD and what are their relevant building blocks.

The hot medium created in heavy-ion collisions is found to be extremely dense and with large cross sections. This leads to interesting transport properties, as a very small viscosity or a large $\hat q$, which are difficult to reconcile with a perturbative approach. Interestingly, this limitations is leading to a flourishing activity on the relation of these findings with theoretically computable quantities in String Theory by the AdS/CFT correspondence which is opening new ways of facing the challenges on the study of collective properties at the fundamental level.

In the next years, the LHC will provide the largest jump in energy in heavy ion collisions ever. With $\sqrt{s}=$5.5 TeV/A these collisions will explore {\it terra incognita} in the phase space diagram of QCD.

\section*{Acknowledgements}
CAS is supported by the FP6  of the European Community under the contract MEIF-CT-2005-024624.

\end{document}